\documentstyle[12pt,epsf,cite]{article}
\newcommand{\C}{{\bf C}}
\renewcommand{\c}[3]{C_{#1#2}{}^{#3}}
\renewcommand{\d}[3]{\Delta_{#1}{}^{#2#3}}
\newcommand{\s}[2]{S^{#1}{}_{#2}}
\def\atuple#1#2#3#4#5#6{\left|
 \begin{array}{@{\,}ccc}
      #1 & #2 & #3 \\ #4 & #5 & #6
 \end{array}\right|}
\def\tuple#1#2#3#4#5#6{\left(
 \begin{array}{@{\,}ccc}
      #1 & #2 & #3 \\ #4 & #5 & #6
 \end{array}\right)}

\begingroup\makeatletter
\def\x#1#2#3#4#5#6#7\relax{\def\x{#1#2#3#4#5#6}}%
\expandafter\x\fmtname xxxxxx\relax %
\gdef\SetFigFont#1#2#3{%
  \ifnum #1<17\tiny\else \ifnum #1<20\small\else
  \ifnum #1<24\normalsize\else \ifnum #1<29\large\else
  \ifnum #1<34\Large\else \ifnum #1<41\LARGE\else
     \huge\fi\fi\fi\fi\fi\fi
  \csname #3\endcsname}%
\makeatother\endgroup
\def\abstract#1{\begin{center}{\large ABSTRACT}\end{center} \par #1}
\def\title#1{\begin{center}{\Large\bf {#1}}\end{center}}
\def\author#1{\begin{center}{\sc #1}\end{center}}
\def\address#1{\begin{center}{\it #1}\end{center}}
\newtheorem{pro}{Proposition}
\newtheorem{lemma}{Lemma}
\newtheorem{theo}{Theorem}
\newtheorem{defi}{Definition}
\begin{document}
\begin{titlepage}
\hspace*{\fill}
\vbox{
  \hbox{TIT/HEP-300}
  \hbox{hep-th/9511044}
  \hbox{November 1995}}
\vspace*{\fill}
\title{On a class of topological quantum field theories in three-dimensions}
\vskip 1cm
\def\thefootnote{\fnsymbol{footnote}}
\author{
  Masako ASANO \footnote{E-mail address: \tt maa@th.phys.titech.ac.jp}} 
\address{
  Department of Physics, 
  Tokyo Institute of Technology, \\
  Oh-Okayama, Meguro,  Tokyo 152, Japan}
\vspace{\fill}
\abstract{
  We investigate the Chung-Fukuma-Shapere theory, 
  or Kuperberg theory, of 
  three-dimensional lattice topological field theory.
  We construct a functor which satisfies the Atiyah's
  axioms of topological quantum field theory by
  reformulating the theory 
  as Turaev-Viro type state-sum theory on a triangulated manifold.
  The theory can also be extended to give a topological invariant 
  of manifolds with boundary.
    } \vspace*{\fill}
\end{titlepage}
\section{Introduction}
Many examples of topological field theory or 
topological invariants 
have been constructed.
Some of them satisfy the axioms of topological
quantum field theory given by Atiyah \cite{at}.
In two dimensions, we know the classification 
of the manifolds well, which leads to the 
complete classification of  
unitary topological quantum field theories on 
compact oriented manifolds~\cite{dj}.
On the other hand, the classification of 
topological field theories of the dimension 
$d\ge 3$ has not been done yet
because of the difficulty of the classification of 
$d$-dimensional manifolds.

In three dimensions, 
it is known that pure gravity theory 
can be interpreted as the 
Chern-Simons-Witten 
theory~\cite{wi88} or the Turaev-Viro theory~\cite{tv}, 
which are both topological theories.
Therefore the investigation of the three-dimensional topological 
field theories for a classification of them 
is important both from mathematical and  
physical point of view.

There are several ways of constructing 
topological field theories in three-dimensions 
such as `surgery' method~\cite{rt,ac},
`state-sum' method~\cite{tv,cfs,ku,dw} and, though it may not be well defined,
the method using functional integral on a manifold $M$~\cite{wi}.

Among many such methods, 
we concentrate on a state-sum model, in particular,
the Chung-Fukuma-Shapere theory~\cite{cfs}.
The Chung-Fukuma-Shapere theory gives an invariant
of closed 3-manifold $M$
for each
involutory Hopf algebra $A$.
It is calculated explicitly by choosing a lattice $L$
of $M$ which is  
a cell complex with some good property. 
Note that the theory is equivalent to the 
invariant given by Kuperberg~\cite{ku} which is defined 
on the basis of triangulations or 
Heegaard diagrams of oriented manifolds.
The theory is well-defined only
for a finite dimensional Hopf algebra $A$
since it suffers divergences if we take an 
infinite dimensional one.
Thus we set $A$ to be finite dimensional. 

\def\thefootnote{\arabic{footnote}}
It seems difficult to extend the
Chung-Fukuma-Shapere invariant 
to a topological quantum field theory satisfying the axioms
of Atiyah in its original form.%
\footnote{
Kuperberg announced in ref.\cite{ku2} that his invariant 
can be extended to give the Atiyah's 
topological quantum field theory.
}
But the situation changes if we 
re-express the invariant as a form similar to  
the Turaev-Viro invariant 
by limiting a lattice to a
simplicial complex, or triangulation.
The Turaev-Viro theory gives the invariant of closed 3-manifolds
and the functor which satisfies Atiyah's axioms.

In this paper,
we explicitly construct the functor of 
topological quantum field theory by defining a
weight, a correspondence of `$q$-$6j$-symbol,'
for a triangulated manifold. 
The method of construction of the functor
we use is similar to that of the Turaev-Viro theory. 

We also show that 
the invariant can be extended to 
that of manifolds with boundary.
It means that we 
give a complex number $\tilde{F}_A(M)$
which is determined only by the topology of $M$
to each compact manifold $M$. 
Our theory gives an extended version of 
Chung-Fukuma-Shapere invariant
whereas the theory 
by Karowski, M\"uller and Schrader
in ref.\cite{kms} gives that of 
the Turaev-Viro invariant.
This fact also shows a similarity 
between the Chung-Fukuma-Shapere invariant and 
the Turaev-Viro invariant.

\section{The Chung-Fukuma-Shapere Theory:
Invariant of Closed 3-manifolds}
Let $M$ be a closed $3$-manifold.
The partition function of the 
Chung-Fukuma-Shapere theory, which is a 
topological invariant of $M$, $Z_{A}(M)$,  is defined 
for each involutory Hopf algebra 
$(A;m,u,\Delta,\epsilon,S)$ over $\C$~\cite{cfs}.
An involutory Hopf algebra is a Hopf algebra with the property that
square of the antipode operator is identity : $S^2= id.$.

We explicitly write the operations on $A$
by means of a basis 
$\{\phi_x|x \in X\}$ of $A$ as follows :
\begin{eqnarray}
  m(\phi_x\otimes\phi_y) & = & \sum_{z\in X}\c{x}{y}{z}\phi_z ,\\
  u(1) & = & \sum_{x\in X}u^x\phi_x,\\
  \Delta(\phi_x) & = & \sum_{y,z\in X}\d{x}{y}{z}\phi_y\otimes\phi_z,\\
  \epsilon(\phi_x) & = & \epsilon_x ,\\
  S(\phi_x) & = & \sum_{y\in X}\s yx \phi_y,
\end{eqnarray}
where  $\c{x}{y}{z}, u^x, \epsilon_x, 1 \in \C$.
Symbols $m$, $u$, $\Delta$, $\epsilon$ and $S$ denote 
multiplication, unit, comultiplication, counit and antipode
respectively.
{}From now on, we often assume that the 
repeated indices are summed over.

We define the metric $g_{xy}$ and the cometric $h^{xy}$ by
\begin{equation}
  g_{xy}\equiv \c xuv \c yvu ,\qquad h^{xy}\equiv \d uvx \d vuy.
\label{metric}
\end{equation}
Since $A$ is involutory, 
$g_{xy}$ and $h^{xy}$ 
have inverses $g^{xy}$ and $h_{xy}$.
We use $g_{xy}$ and $g^{xy}$ to raise and lower the indices of 
$\c xyz$ and $u^x$ (e.g., $C_{xyz}=g_{zu}\c xyu$).
Similarly, we use $h^{xy}$ and $h_{xy}$ for 
$\d xyz$ and $\epsilon_x$.

We summarize some important relations 
among $C$, $\Delta$ and $S$ which hold generally for
an involutory Hopf algebra.
Some of these relations play an important role in verifying the 
topological invariance of $Z_A(M)$.
\begin{eqnarray}
 && C_{x_1x_2\cdots x_n} = C_{x_nx_1\cdots x_{n-1}}\label{cyclc}\, ,\\
 && \Delta^{x_1x_2\cdots x_n} = \Delta^{x_nx_1\cdots x_{n-1}}\, ,\\
 && S^{x}{}_{y}= |X|^{-1}g^{xz}h_{zy}\label{sgh}\, ,\\
 &&  |X|^{-1}C_{x_1x_2\cdots x_n}\, C_{y_1y_2\cdots y_n}\, 
   \Delta_{z_1}{}^{x_1y_1}\,\Delta_{z_2}{}^{x_2y_2}\,
   \cdots \,\Delta_{z_n}{}^{x_ny_n}
   =C_{z_1z_2\cdots z_n}\, ,
\label{ccddd}\\
  &&C_{x_1x_2\cdots x_n}\,S^{x_1}{}_{y_1}\,S^{x_2}{}_{y_2}\,\cdots 
  \,S^{x_n}{}_{y_n} =
  C_{y_n\cdots y_2y_1}
\label{csss}\\
  &&\Delta^{x_1x_2\cdots x_n}\,S^{y_1}{}_{x_1}\,S^{y_2}{}_{x_2}\,\cdots 
  \,S^{y_n}{}_{x_n} =
  \Delta^{y_n\cdots y_2y_1}
\end{eqnarray}
where $|X|$ is an order of the algebra $A$ and 
\begin{eqnarray}
    C_{x_1x_2\cdots x_n} &\equiv &
    \c{a_1}{x_1}{a_2} \c{a_2}{x_2}{a_3} \times \cdots
    \times \c{a_{n-1}}{x_{n-1}}{a_n} \c{a_n}{x_n}{a_1} \, ,\\
    \Delta^{x_1x_2\cdots x_n}
    &\equiv & \d {a_1}{x_1}{a_2}\d
    {a_2}{x_2}{a_3}\times\cdots\times  
    \d {a_{n-1}}{x_{n-1}}{a_{n}}
    \d {a_{n}}{x_n}{a_1} .   
\end{eqnarray}

With these preparations, 
we now recall the 
definition of the invariant $Z_{A}(M)$ 
given in ref.\cite{cfs}.

We first choose a lattice $L$ which
represents $M$. 
Here a lattice $L$ is a three-dimensional cell complex 
such that every $2$-cell is a polygon and 
every 1-cell is a boundary of at least three $2$-cells. 
The definition is given as follows :
\begin{enumerate}
\item
  Decompose $L$ into the set of polygonal faces $F=\{f_i\}_{i=1,...,N_2}$
  and that of hinges $H=\{h_i\}_{i=1,...,N_1}$ as depicted in fig.~\ref{fahi}.
  Here $N_i$ denotes the number of $i$-cells in $L$.
\begin{figure}[bthp]
    \begin{center}
      \leavevmode
       \epsfysize = 5.5cm 
       \epsfbox{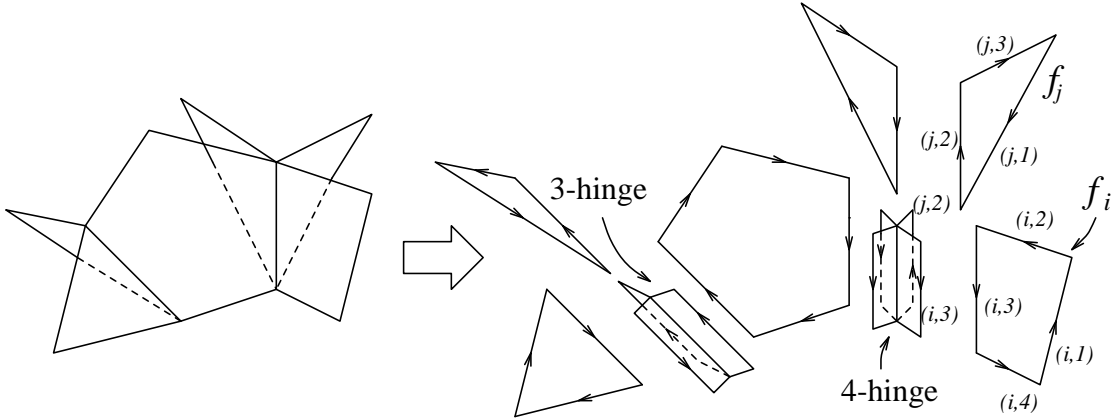}
    \end{center}
    \caption{The decomposition of a part of a lattice $L$
     into faces and hinges.
      An $n$-hinge pastes $n$ different faces.
     Arrows are laid on each edge of faces or hinges.}     
\label{fahi}
\end{figure}
  We pick an orientation of each face $f_i$ and put an arrow on each 
  edge according to it.  We associate 
  symbols $(i,1),(i,2),...,(i,n)$ to the edges of each $n$-gon $f_i$
  (Fig.\ref{fahi}).
\item
Assign
$C_{x_{(i,1)}x_{(i,2)}\ldots x_{(i,n)}} \in \C$
  to each $n$-gonal face $f_i$
where the index $x_{(i,j)}$ is an element of $X$. 
\item
  The assignment of an arrow and a symbol to each edge of faces induces
  those to each edge of hinges $\{h_i\}$. 
  The $m$ arrows on the edges of a $m$-hinge $h$ are not always in
  the same direction. 

  If all arrows in the $m$-hinge $h$ are in the same direction,
  we align the indices on the edges of the hinge associated as above
  in the clockwise
  order of the edges around the arrows 
  as $(i_1,j_1),\,(i_2,j_2),\,\cdots\, (i_m,j_m)$
  and associate
  $    \Delta^{x_{(i_1,j_1)}x_{(i_2,j_2)}\cdots x_{(i_m,j_m)}} \in \C$
  to $h$.  

  If directions of arrows on some edges of  $h$ are not the same as the rest,
  we change the direction of the arrows so as to make directions of
  all the arrows match by multiplying an
  additional factor (the direction changing operators) 
  $\s{x}{x'} \in \C $ (See fig.~\ref{dich}) for each edge
  of which we would like to upside down the direction of arrow. 
\begin{figure}
    \begin{center}
      \leavevmode
      \epsfysize=3.5cm
      \epsfbox{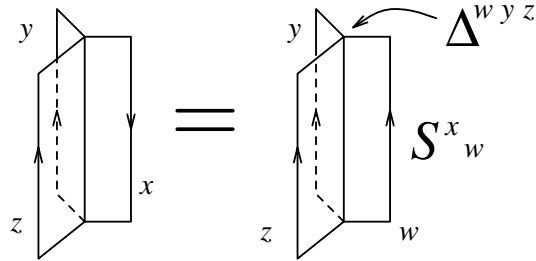}
    \end{center}
    \caption{The operation of the direction changing operator $S$.
      The role of the hinge in the left-hand side is 
      the same as that in the right-hand side : 
      $\Delta^{wyz}\times S^{x}{}_{w}$ .} 
    \label{dich}
\end{figure}
\item
  So far we have defined the weight 
  $C_{x_{(i,1)}\cdots x_{(i,n)}}$ for each face $f_i$ and \\
  $\Delta^{x_{(i_1,j_1)}\cdots x'_{j_k} \cdots x_{(i_m,j_m)}} 
  \prod_{j_k\in R_h} \s{x_{(i_k,j_k)}}{x'_{j_k}}$
  for each hinge $h$. 
  The set $R_h$ corresponds to the set of all direction changing
  edges of a hinge $h$.
  The partition function is defined by contracting indices as
  \begin{equation}
    Z_{A}(L)
    =
    |X|^{-N_1-N_3}
    \prod_{i=1}^{N_2}  C_{x_{(i,1)}\cdots x_{(i,n_i)}}
    \prod_{h\in H} 
    \left[\Delta^{x_{(i_1,j_1)}\cdots x'_{j_k} \cdots x_{(i_m,j_m)}}
    \prod_{j_k \in R_h} \s{x_{(i_k,j_k)}}{x'_{j_k}} \right]\, .
    \label{zfuk}
  \end{equation}
\end{enumerate}

It is shown that this value does not depend on 
the direction of arrows on edges of each face
nor any local deformation of a lattice $L$ 
which preserves a topology of $L$.
Thus, 
\begin{theo}[Chung-Fukuma-Shapere~\cite{cfs}]
  $Z_{A}(L)$ is a topological invariant of a manifold $M$.
\label{invcfs}
\end{theo}

Note that
in particular, if a lattice $L$ is a simplicial complex, 
$Z_{A}(L)$ is invariant under Alexander moves of 
$L$\cite{al}.

\section{Some Examples}
We calculate $Z_A$ for some manifolds $M$.

\begin{figure}[bthp]
    \begin{center}
      \leavevmode
       \epsfysize = 3.5cm 
       \epsfbox{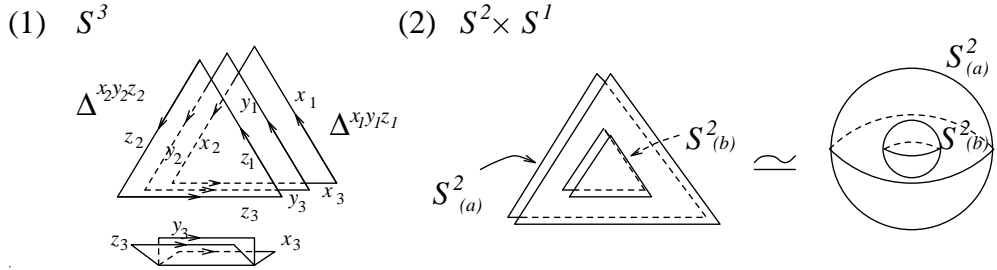}
    \end{center}
    \caption{(1) $S^3$ : $N_3=N_1=3$ ;  
             (2) $S^2\times S^1$: Outside of the sphere $S^2{}_{(a)}$
         and  inside of $S^2{}_{(b)}$ are identified. 
         $N_3=N_1=3$.}
\label{example}
\end{figure}
For a three-sphere $S^3$, we can take a lattice 
consisting of three triangles and three hinges (Fig.\ref{example}(1)).
The invariant $Z_A$ is calculated as 
\begin{eqnarray}
Z_A(S^3) &=& |X|^{-3-3}\,
C_{x_1x_2x_3}C_{y_1y_2y_3}C_{z_1z_2z_3}
\Delta^{{x_1}{y_1}{z_1}} \Delta^{{x_2}{y_2}{z_2}}
\Delta^{{x_3}{y_3}{z_3}}
\nonumber\\
&=& |X|^{-5}\,C_{w_1w_2w_3}C_{z_1z_2z_3}
h^{z_1w_1}h^{z_2w_2}h^{z_3w_3}
\nonumber\\
&=& |X|^{-2}\,C_{z_1z_2z_3}C_{u_3u_2u_1}g^{z_1u_1}g^{z_2u_2}g^{z_3u_3}
\nonumber\\
&=& |X|^{-2}\,C_{z_1z_2}{}^{u_3}C_{u_1u_3}{}^{z_2}g^{z_1u_1}
\nonumber\\
&=& |X|^{-2}\,g_{z_1u_1}g^{z_1u_1}
\nonumber\\
&=& |X|^{-1}
\label{zas3}
\end{eqnarray}
where we use the relations (\ref{metric}), (\ref{sgh}),
(\ref{ccddd}) and (\ref{csss}).

The next example is a manifold $S^2\times S^1$.
Considering the lattice in Fig.\ref{example}(2), 
the invariant can be evaluated as 
\begin{eqnarray}
Z_A(S^2\times S^1)&=& |X|^{-3-3}
C_{x_1x_2x_3}C_{y_1y_2y_3}C_{z_1z_2z_3}C_{w_1w_2w_3}
\Delta^{x_1y_1z_1w_1}\Delta^{x_2y_2z_2w_2}\Delta^{x_3y_3z_3w_3}
\nonumber\\
&=& 
|X|^{-4}C_{u_1u_2u_3}C_{v_1v_2v_3}h^{u_1v_1}h^{u_2v_2}h^{u_3v_3}
\nonumber\\
&=& 1.
\label{zas2s1}
\end{eqnarray}
Here we use the relation 
$\Delta^{x_1y_1z_1w_1}=\d {u_1}{x_1}{y_1} \d {v_1}{z_1}{w_1} h^{u_1v_1}$.

Note that the values $Z_A(S^3)$ and
$Z_A(S^2\times S^1)$ are both independent of the choice of the 
algebra $A$.

\section{Another Representation}\label{secfm}
In this section, we give another representation of 
the Chung-Fukuma-Shapere invariant.
The idea is to take a simplicial complex, or a triangulation,
as a lattice $L$ and
define 
a weight $W_i$, 
which acts similarly as a quantum $6j$-symbol in the case of 
the Turaev-Viro theory,
  on each tetrahedron $T_i$.
This makes the invariant $Z_A(L)$ the form
\begin{equation}
   Z_A(L) = N \sum\prod_{i=1}^{N_3} W_i
\label{z1}
\end{equation}
where $N$ is a normalization factor which will be 
given explicitly later.

\begin{figure}[tbhp]
    \begin{center}
      \leavevmode
      \epsfxsize=16cm
      \epsfbox{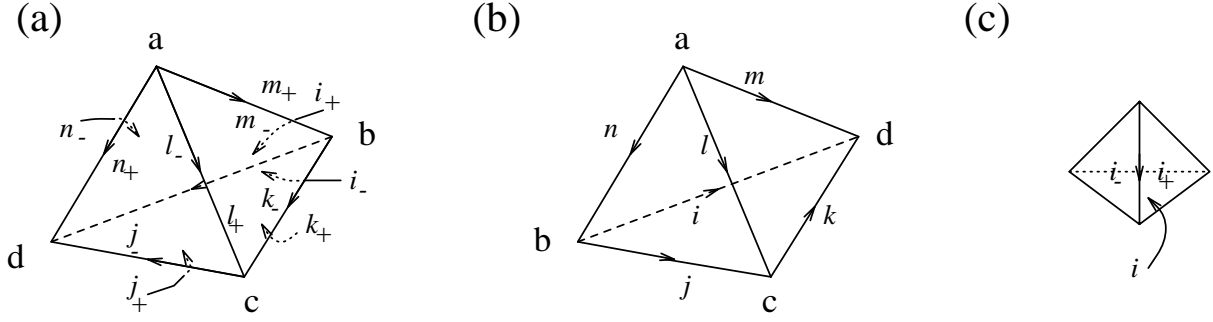}
    \end{center}
    \caption{Tetrahedron of class (a) $U^+$ or (b) $U^-$,
     where $a>b>c>d$.
    We use the simplified notation in the figure (b). The rule 
    is as same as (a), which is denoted in (c).} 
    \label{tetra+-}
\end{figure}

Now we begin by making preparations for obtaining the 
form eq.(\ref{z1}) from eq.(\ref{zfuk}).
Let
$M$ be a triangulated manifold, i.e., a simplicial complex
representing a certain manifold.
We number all the vertices of $M$ arbitrarily as $1,2,...,N_0$,
and according to that we associate an arrow along each $1$-simplex (edge)
as follows : for an edge whose boundary 
consists of vertices $i$ and $j$,
the direction of an arrow on the edge is  $i\longrightarrow j$ if $i>j$,
and $i\longleftarrow j$ otherwise.%
\footnote{We choose such a way of defining 
the direction of arrows
only for simplicity.
In practice, we can define $Z_A(L)$ of (\ref{z1})
for a manifold whose arrows on 1-simplices
are given arbitrarily, though it
is rather complicated.
We comment on the point later again.} 
In this way, 
all tetrahedra $T_i$ ($i\in \{1,2,...,N_3\}$)
are divided into two classes, $U^+$ and $U^-$,
according to the orientation of the order of four vertices $a$,
$b$, $c$ and $d$ of $T_i$.
We define the tetrahedra whose vertices are oriented as 
(a) (or (b)) of Fig.\ref{tetra+-} 
to be in a class $U^+$ (or class $U^-$) 
under the assumption that $a>b>c>d$.

Next, we `color' the triangulated manifold $M$ by 
associating a symbol which is an element of $X$
to each pair of a $1$-simplex $E_i$ and a $2$-simplex $F_j$ in $M$
if $E_i \cap F_j = E_i$.
This gives $3\times N_2$ symbols on $M$.
To be concrete, a coloring $\phi$ on $M$ is a map
\begin{equation}
  \phi : \{(E_i, F_j) \, |\, i=1,\cdots,N_1 ,\, j=1,\cdots, N_2 ,\,
     E_i\cap F_j =E_i \}\longrightarrow X.
  \label{colorphi}
\end{equation}
Note that a coloring on $M$ induces a coloring 
on each tetrahedron $T_i$ (Fig.\ref{tetra+-}).

Then, given an involutory Hopf algebra,
we define a weight $W_i{}^{\kappa_i}$
on a tetrahedron $T_i$  which is given coloring and 
arrows on edges as Fig.\ref{tetra+-} (a) or (b) :
\begin{equation}
  W_i{}^{\kappa_i}=\left\{
    \begin{array}{ll}
      W_i{}^+ & \quad {\rm for} \,\,\,\, T_i \in U^+\\    
      W_i{}^- & \quad {\rm for} \,\,\,\, T_i \in U^-    
    \end{array}
  \right.
\end{equation}
where
\begin{eqnarray}
  W_i{}^+ \tuple {i_+,i_-} {j_+,j_-} {k_+,k_-} {l_+,l_-} 
      {m_+,m_-} {n_+,n_-} &=& 
   C_{{I'}KJ}\, C_{{K'}{M'}L}\, C_{{J'}{L'}N}\, C_{I{N'}M} 
   \,S^{I'}{}_{I''}\,\Delta^{I''Ii_+}{}_{i_-}\,
   \nonumber\\
   & & \hspace{-1cm}\times 
       S^{J'}{}_{J''}\,\Delta^{J''Jj_+}{}_{j_-}\,
       S^{K'}{}_{K''}\,\Delta^{K''Kk_+}{}_{k_-}\,
       S^{L'}{}_{L''}\,\Delta^{L''Ll_+}{}_{l_-}\,\nonumber\\
   & & \hspace{-1cm}\times 
       S^{M'}{}_{M''}\,\Delta^{M''Mm_+}{}_{m_-}\,
       S^{N'}{}_{N''}\,\Delta^{N''Nn_+}{}_{n_-}
\label{w+}
\end{eqnarray}
and
\begin{eqnarray}
  W_i{}^{-}\tuple {i_+,i_-} {j_+,j_-} {k_+,k_-} {l_+,l_-} 
      {m_+,m_-} {n_+,n_-} &=&
     C_{{I}{K'}{J'}}\, C_{{K}{M'}L}\, C_{{J}{L'}N}\, C_{{I'}{N'}M}\,
     S^{I'}{}_{I''}\,\Delta^{I''Ii_+}{}_{i_-}\, \nonumber\\
   & & \hspace{-1cm}\times 
       S^{J'}{}_{J''}\,\Delta^{J''Jj_+}{}_{j_-}\,
       S^{K'}{}_{K''}\,\Delta^{K''Kk_+}{}_{k_-}\,
       S^{L'}{}_{L''}\,\Delta^{L''Ll_+}{}_{l_-}\,\nonumber\\
   & & \hspace{-1cm}\times 
       S^{M'}{}_{M''}\,\Delta^{M''Mm_+}{}_{m_-}\,
       S^{N'}{}_{N''}\,\Delta^{N''Nn_+}{}_{n_-}.
\label{w-}
\end{eqnarray}
The indices $I,I',I'',J,... $ are elements of a set $X$
and summed over in these equations.
We define 
\begin{equation}
  F(M)\equiv |X|^{-2N_2-N_0}\sum_{\{\phi\}}\prod_{i=1}^{N_3}W_i{}^{\kappa_i}
  \in \C
  \label{ft}
\end{equation} 
where sum of $\phi$ is taken over all the maps 
satisfying (\ref{colorphi}).
\begin{pro}
  $F(M) = Z_A(M)$ .
\label{fza}
\end{pro}
{\it Proof.}     

First, we explicitly write down sum over colorings $\{\phi\}$ on $M$
in eq.(\ref{ft}).
We pay attention to a $1$-simplex $E_i$ and give 
numbers $1,2,...,m_i$
to all triangles which include $E_i$ as an edge 
by a clockwise order with respect to the direction 
of the arrow on $E_i$.
This induces a color $x^i_{k}$ 
on a pair $(E_i , F_{k})$ 
where $k=1,\cdots m_i$.
Denoting all colorings of $M$ as the same way, 
we rewrite the sum over all colorings $\{\phi\}$ as 
\begin{equation}
  \sum_{\{\phi\}}\rightarrow 
  \prod_{i=1}^{N_1}\sum_{x^i_1,x^i_2,\cdots x^i_{m_i}\in X} .
\end{equation}
Extracting a part which depends on the sum 
of $x^i_1,x^i_2,\cdots x^i_{m_i}\in X$
from eq.(\ref{ft}) 
by use of (\ref{w+}) and (\ref{w-}),
and performing the calculation, we get for each 
1-simplex $i$ 
\begin{equation}
   \sum_{x^i_1,x^i_2,\cdots x^i_{m_i}\in X}
    \Delta^{z^i_1 y^i_1 x^i_2}{}_{x^i_1}
    \Delta^{z^i_2 y^i_2 x^i_3}{}_{x^i_2}\cdots
    \Delta^{z^i_{m_i} y^i_{m_i} x^i_1}{}_{x^i_{m_i}}
   = \Delta^{z^i_1\, y^i_1\, z^i_2\, y^i_2\cdots z^i_{m_i}\,y^i_{m_i}} .
\end{equation}
Thus eq.(\ref{ft}) is rewritten as 
\begin{eqnarray}
   F(M)&=&|X|^{-N_1-N_2-N_3}
   \prod_{i=1}^{N_1}\sum_{z^i_l,y^i_l,y^{i+N_1}_l}
   \left[\Delta^{z^i_1 y^i_1 z^i_2 y^i_2\cdots z^i_{m_i}y^i_{m_i}}\,
       S^{y^{i+N_1}_1}{}_{z^i_1}\,S^{y^{i+N_1}_2}{}_{z^i_2}\,\cdots
       S^{y^{i+N_1}_{m_i}}{}_{z^i_{m_i}}
   \right] \nonumber\\ 
   &&\times \prod_{k=1}^{N_2} C_{y^{j_1}_{k_1}y^{j_2}_{k_2}\,y^{j_3}_{k_3}}
         C_{y^{j_{\alpha_1}}_{k_{\alpha_1}}\,y^{j_{\alpha_2}}_{k_{\alpha_2}}
         y^{j_{\alpha_3}}_{k_{\alpha_3}}}.
\label{ft2}
\end{eqnarray}
Here we use the Poincare duality theorem 
$N_3-N_2+N_1-N_0=0$.
The quantity
$C_{y^{j_1}_{k_1}y^{j_2}_{k_2}\,y^{j_3}_{k_3}}\times$ 
$C_{y^{j_{\alpha_1}}_{k_{\alpha_1}}\,y^{j_{\alpha_2}}_{k_{\alpha_2}}
 y^{j_{\alpha_3}}_{k_{\alpha_3}}}$
comes from two tetrahedra 
whose intersection is 
 a triangle $F_k$ and 
$(k_{\alpha_1},k_{\alpha_2},k_{\alpha_3})$ is a permutation of 
$(k_1,k_2,k_3)$.
Thus
we see that the value (\ref{ft2}) is the 
Chung-Fukuma-Shapere invariant $Z_A(L)$ 
for a lattice $L$ which is generated by gluing 
$N_3$ simplices $\{T_i \}_{i=1,\cdots N_3}$.
Note that $L$ is a lattice $M$
whose faces are all duplicated.
Thus it 
is different from  
$M$ as a lattice and it has
$N_3+N_2$ 3-cells and  $2\times N_2$ 2-cells.
Since $Z_A(L)$ is a topological invariant, it is the same as $Z_A(M)$.
Thus the proof is completed.
\rule[-1pt]{6pt}{6pt}
\medskip

Note that 
for a triangulated manifold $M$ with arrows of arbitrary direction 
not induced from the order of the vertices,  
$F(M)$ can also be defined.
In this case, since it is possible that there exists 
a 3-simplex $T$ which belongs to neither $U^+$ nor $U^-$,
we have to give a definition of $W{}^{\kappa}$
for such $T$.  
For example, for a 3-simplex $T$
whose arrows and weights are the same as $T_+\in U^+$
except the arrow on an edge $j$ is in the opposite direction, 
the weight is obtained by 
multiplying $S\cdot S$ as follows :
\begin{equation}
 W^{\kappa} \tuple {i_+,i_-} {j_+,j_-} {k_+,k_-} {l_+,l_-} 
      {m_+,m_-} {n_+,n_-} 
  =  \sum_{j'_+,j'_-\in X}W^+ \tuple {i_+,i_-} {j'_-,j'_+} 
        {k_+,k_-} {l_+,l_-} {m_+,m_-} {n_+,n_-}  
      S_{j_-\,j'_-}\,S^{j_+\,j'_+} .
\label{wws}
\end{equation}
The weight for any other 3-simplex can be 
obtained similarly.
For simplicity, we write eq.(\ref{wws}) as 
\begin{equation}
 W^\kappa \tuple {\bar{i}_\uparrow} {\bar{j}_\uparrow} {\bar{k}_\uparrow} 
   {\bar{l}_\uparrow} {\bar{m}_\uparrow} {\bar{n}_\uparrow} 
  = W^+ \tuple {\bar{i}_\uparrow} {\bar{j}'_\downarrow} {\bar{k}_\uparrow} 
   {\bar{l}_\uparrow} {\bar{m}_\uparrow} {\bar{n}_\uparrow} 
      \bar{S}^{\bar{j}}{}_{\bar{j}'}
\end{equation}
where the new symbols $\bar{i}_\uparrow$ and $\bar{i}_\downarrow$ 
stand for 
pairs
$(i_+,i_-)$ and $(i_-,i_+)$ respectively and  
\begin{equation}
   \bar{S}^{\bar{j}}{}_{\bar{j}'} = S_{j_-\,j'_-}\,S^{j_+\,j'_+} . 
\end{equation}
We can consider that $\bar{i}_\uparrow$ and $\bar{i}_\downarrow$ 
are the same colored 1-simplex but the directions of arrows 
are opposite to each other.

\bigskip

Now we give some properties of the weight $W{}^{\kappa}$ for a 
tetrahedron $T$.

\begin{figure}
    \begin{center}
      \leavevmode
      \epsfxsize=14cm
      \epsfbox{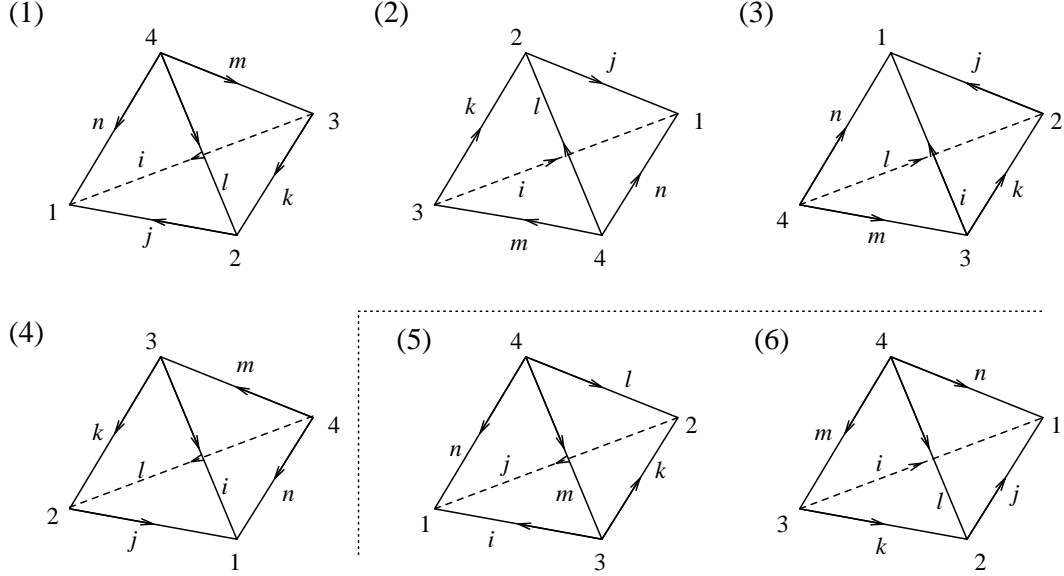}
    \end{center}
    \caption{Symmetries of the weight $W$: (1)-(4) 
    are the colored tetrahedra with the same orientation
  and others are those with the opposite orientation. 
 Arrows are associated such that the tetrahedron (1) is the same as the (a)
of Fig.\protect\ref{tetra+-}.}  
   \label{symmetry}
\end{figure}

Remember that 
in the case of the Turaev-Viro theory
the quantum $6j$-symbol $|:::|$ %
\footnote{The theory can also be defined by using other `symbol'
than quantum $6j$-symbol 
which satisfies a certain property.\cite{tv}}
defined for a tetrahedron with admissible color on six edges of it
has the symmetries 
\begin{equation}
  \atuple i j k l m n =   
  \atuple i m n l j k =   
  \atuple l m k i j n =   
  \atuple l j n i m k =   
  \atuple j i k m l n =   
  \atuple i k j l n m  .
\label{sym6j}
\end{equation}
These symmetries correspond to the 
rotational symmetry and orientation changing 
symmetry of the tetrahedron,
 namely the symmetries 
among the six colored tetrahedra depicted in Fig.\ref{symmetry}.
Here we use the term orientation as 
that of the tetrahedron as a 2-sphere $S^2$.

In our case of the weight $W{}^{\kappa}$, 
the relation (\ref{sym6j}) is not satisfied in its original form since
each 1-simplex has an arrow in addition to the color.
Instead, we have a modified version.
In the case of a tetrahedron $T$ of the class 
$U^+$ or $U^-$,
it can be written as follows:
\begin{eqnarray}
& & \hspace{-7mm}
W^\pm \tuple {\bar{i}_\uparrow} {\bar{j}_\uparrow} {\bar{k}_\uparrow} 
   {\bar{l}_\uparrow} {\bar{m}_\uparrow} {\bar{n}_\uparrow}
\label{31} \\ 
& =& \sum_{\bar{i}'\bar{k}'\bar{l}'\bar{n}'}
 W^\pm \tuple{\bar{i}'_\downarrow} {\bar{m}_\uparrow} {\bar{n}'_\downarrow} 
   {\bar{l}'_\downarrow} {\bar{j}_\uparrow} {\bar{k}'_\downarrow} 
      \bar{S}^{\bar{i}}{}_{\bar{i}'}\bar{S}^{\bar{k}}{}_{\bar{k}'}
      \bar{S}^{\bar{l}}{}_{\bar{l}'}\bar{S}^{\bar{n}}{}_{\bar{n}'}\\
& =& \sum_{\bar{i}'\bar{j}'\bar{k}'\bar{l}'\bar{m}'\bar{n}'}
 W^\pm\tuple{\bar{l}'_\downarrow}
     {\bar{m}'_\downarrow} {\bar{k}'_\downarrow} 
   {\bar{i}'_\downarrow} {\bar{j}'_\downarrow} {\bar{n}'_\downarrow} 
      \bar{S}^{\bar{i}}{}_{\bar{i}'} \bar{S}^{\bar{j}}{}_{\bar{j}'}
      \bar{S}^{\bar{k}}{}_{\bar{k}'}\bar{S}^{\bar{l}}{}_{\bar{l}'}
      \bar{S}^{\bar{m}}{}_{\bar{m}'}\bar{S}^{\bar{n}}{}_{\bar{n}'}\\
& =&\sum_{\bar{j}'\bar{m}'}
 W^\pm \tuple {\bar{l}_\uparrow} {\bar{j}'_\downarrow} {\bar{n}_\uparrow} 
   {\bar{i}_\uparrow} {\bar{m}'_\downarrow} {\bar{k}_\uparrow} 
      \bar{S}^{\bar{j}}{}_{\bar{j}'}\bar{S}^{\bar{m}}{}_{\bar{m}'} .
\label{34}  
\end{eqnarray}
where eqs.(\ref{31})-(\ref{34})  correspond to  the
tetrahedra (1)-(4) in Fig.\ref{symmetry} respectively.
Note that the above relations are only among tetrahedra with
the same orientation.
Generically, there is no corresponding symmetry between 
two tetrahedra of different orientation. 

The weight also obeys the relation   
\begin{eqnarray}
&&\hspace{-5mm}
 W^+ \tuple {i_+,i} {j,j_-} {k,k_-} {l_+,l_-} 
      {m_+,m_-} {n_+,n_-} 
 W^+ \tuple {k_-,k'} {j_{0+},j_{0-}} {i',i_+} {n_-,n_+} 
      {m_-,m_+} {l_-,l_+} 
 S_{j_{0+} j'} S^{j_{0-} j_-}
\nonumber\\
&=& 
 W^+ \tuple {i_+,i} {j,j_-} {k,k_-} {l_+,l_-} 
      {m_+,m_-} {n_+,n_-} 
 W^- \tuple {i',i_+} {k_-,k'} {j_-,j'} 
{l_-,l_+}{n_-,n_+} {m_-,m_+} 
\nonumber\\
&=& 
|X|^{8} C_{IKJ}\Delta^{Ii_0}{}_{i_0'} S_{i_0 i} S^{i_0' i'}
\Delta^{Kk}{}_{k'}\Delta^{Jj}{}_{j'}.
\label{bubble}
\end{eqnarray}
\begin{figure}
    \begin{center}
      \leavevmode
      \epsfxsize=16cm
      \epsfbox{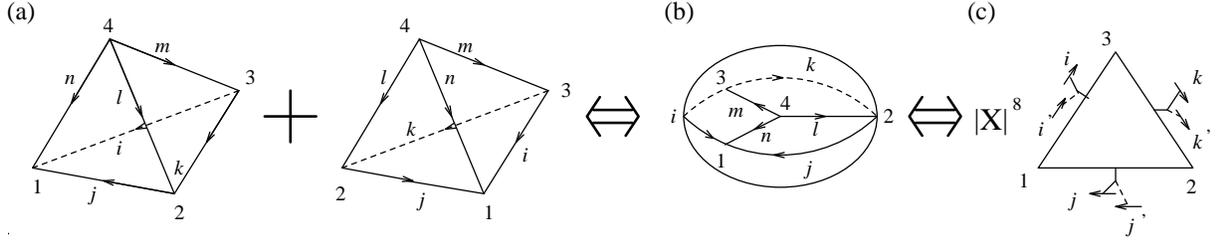}
    \end{center}
    \caption{The relation (\protect\ref{bubble}):
    Connection of two tetrahedra in (a) by three edges
    $l$, $m$ and $n$ induces (b), which gives the same weight as that of 
    (c).}
   \label{bubfig}
\end{figure}
{}From this equation we obtain the orthogonality relation 
analogous to that of quantum $6j$-symbol case :
\begin{eqnarray}
&&\hspace{-5mm} 
 W^+ \tuple {i_+,i_-} {j,j_-} {k_+,k_-} {l_+,l_-} 
      {m_+,m_-} {n_+,n_-} 
 W^+ \tuple {k_-,k_+} {j_{0+},j_{0-}} {i_-,i_+} {n_-,n_+} 
      {m_-,m_+} {l_-,l_+} 
 S_{j_{0+} j'} S^{j_{0-} j_-}\nonumber\\
&=& 
 W^+ \tuple {i_+,i_+} {j,j_-} {k_+,k_-} {l_+,l_-} 
      {m_+,m_-} {n_+,n_-} 
 W^- \tuple {i_-,i_+} {k_-,k_+} {j_-,j'} 
{l_-,l_+}{n_-,n_+} {m_-,m_+} 
\nonumber\\
 & =&|X|^{10} \delta^{j}{}_{j'}.
\label{orthogonal}
\end{eqnarray}
We also have analogous relations to 
the Biedenharn-Elliot identities of the 
quantum $6j$-symbols :
\begin{eqnarray}
&&\hspace{-5mm} 
|X|^{-4}
 W^+ \tuple {n_1,n_2} {j_3,j_2} {l_3,l_2} {q_1,q_2} 
      {p_1,p_3} {r_1,r_3} 
\nonumber\\
&&\hspace{1cm}\times
 W^+ \tuple {i_3,i_2} {j_1,j_3} {k_1,k_2} {l_1,l_3} 
      {m_3,m_2} {n_2,n_3} 
 W^- \tuple {n_3,n_1} {m_1,m_3} {i_1,i_3} {s_1,s_2} 
      {r_3,r_2} {p_3,p_2} 
\nonumber\\
&=& 
 W^+ \tuple {i_1,i_2} {j_1,j_2} {k_1,k_3} {q_3,q_2} 
      {s_1,s_3} {r_1,r_2} 
 W^- \tuple {l_1,l_2} {m_1,m_2} {k_3,k_2} {s_3,s_2} 
      {q_1,q_3} {p_1,p_2} \quad .
\label{32move}
\end{eqnarray}
\begin{figure}
    \begin{center}
      \leavevmode
      \epsfxsize=7cm
      \epsfbox{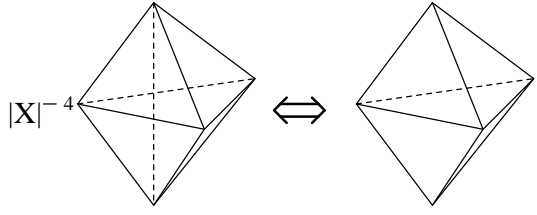}
    \end{center}
    \caption{The relation (\protect\ref{32move}). It corresponds to 
   the topology preserving move, $(3,2)$-move.}
   \label{biefig}
\end{figure}
By changing the direction of arrows on some 1-simplices by multiplying 
$S\cdot S$,
we obtain an
equation of another type, for example,
\begin{eqnarray}
&&\hspace{-5mm} |X|^{-4}
 W^+ \tuple {t_3,t_2} {u_1,u_3} {j_1,j_2} 
            {n_1,n_3} {l_3,l_2} {r_1,r_2} 
\nonumber\\
&&\hspace{1cm}\times
 W^+ \tuple {s_3,s_2} {t_1,t_3} {k_1,k_2} 
            {l_1,l_3} {m_3,m_2} {r_2,r_3} 
 W^- \tuple {s_1,s_3} {i_1,i_2} {u_3,u_2} 
            {n_3,n_2} {r_3,r_1} {m_1,m_3} 
\nonumber\\
&=& 
 W^+ \tuple {i_3,i_2} {j_3,j_2} {k_3,k_2} 
            {l_1,l_2} {m_1,m_2} {n_1,n_2} 
 W^+ \tuple {t_1,t_2} {u_1,u_2} {j_1,j_3} 
            {i_1,i_3} {k_1,k_3} {s_1,s_2} \quad .
\label{32move2}
\end{eqnarray}

\section{Manifolds with Boundary}
{}From now on,
we consider an compact triangulated
3-manifold $M$ with 
an arrow along each 1-simplex. 
We denote $M_i$ $(i=0,1,2)$ the number of $i$-simplices 
of the boundary of $M$ : $\partial M$.  

Let $\phi$ be a coloring of $M$ as (\ref{colorphi}).
In particular, we call
$\partial \phi$ a coloring of $\partial M$ :
\begin{equation}
  \partial \phi : \{(E_i, F_j) \, |\, E_i, F_j\in \partial M\, ,
     E_i\cap F_j =E_i \}\longrightarrow X.
\end{equation}
For a fixed coloring $\partial \phi$ on the boundary $\partial M$
and an involutory Hopf algebra $A$, 
we define 
\begin{equation}
  F(M;\partial \phi) = |X|^{-2 N_2-N_0+(2M_2+M_0)/2}
    \sum_{\{\phi\}|_{\partial \phi}} 
    \prod_{i=1}^{N_3}W_i{}^{\kappa_i}\in \C .
\label{fmphi}
\end{equation}
Here we denote $\{\phi\}|_{\partial \phi}$ 
the set of all colorings $\{\phi\}$ with a fixed coloring 
$\partial \phi$ on $\partial M$. 
Note that in the case of a closed manifold,
eq.(\ref{fmphi}) reduces to the Chung-Fukuma-Shapere
invariant eq.(\ref{zfuk}) or eq.(\ref{ft}). 
\begin{pro}
 $F(M;\partial \phi)$ does not depend on the 
 direction of an arrow on a 1-simplex in $ M\setminus\partial M$ and 
 it is invariant under local topology preserving deformation 
 of $M\setminus\partial M$. 
\label{invariantd}
\end{pro}
The proof of this proposition is straightforward from 
proposition~\ref{fza} and the 
note under the theorem~\ref{invcfs}
since we know that all topology preserving deformation of 
$M\setminus\partial M$
is generated by a finite sequence of Alexander moves 
and their inverses.

\section{Construction of the Functor}\label{functor}
In this section, we interpret  $F(M;\partial \phi)$ as 
the operator on a boundary of $M$ and
construct a functor which satisfies Atiyah's axioms  
of 3-dimensional topological quantum field theory\cite{at}.
The construction can be done along lines similar to ref.\cite{tv}.

A 3-dimensional cobordism $(M, \Sigma_1, \Sigma_2)$ is an
compact 3-dimensional manifold $M$ 
together with 
two closed oriented surfaces $\Sigma_1$, $\Sigma_2$, such that
\begin{equation}
  \Sigma_1\cap \Sigma_2 =\phi,
  \quad
  \partial M = (-\Sigma_1)\cup(\Sigma_2).
\end{equation}
Note that it induces a category whose objects and morphisms
are closed surfaces and 3-dimensional cobordisms
respectively.

For a given cobordism $(M, \Sigma_1, \Sigma_2)$,
we assume that 
the manifold $M$ is triangulated and is given an arrow along each 1-simplex,
which induces a triangulation and arrows on $\Sigma_1$ and $\Sigma_2$.
Then we define a $\C$-module $V_{\Sigma_i}$,
which is freely generated by colorings of $\Sigma_i$,
for a triangulated surface $\Sigma_i$ $(i=1,2)$
with arrows on all 1-simplices.
The dimension of $V_{\Sigma_i}$ is 
\begin{equation}
\dim \, V_{\Sigma_i} = 2\, l^i\, |X|
\end{equation}
where $l^i$ is the number of 1-simplices in $\Sigma_i$. 
If $\Sigma=\phi$ then we set $V_{\Sigma_i}=\C$.
{}From now on, we use the symbol $\Sigma_i$ as 
a triangulated surface equipped with arrows 
in the sense we described in previous sections.

We define a homomorphism 
from $V_{\Sigma_1}$ to $V_{\Sigma_2}$ by  
\begin{equation}
F_{12}(\alpha_1)\,=\,\sum_{\alpha_2}F(M;\alpha_1\cup\alpha_2)\,\alpha_2
\qquad : V_{\Sigma_1}\rightarrow V_{\Sigma_2}
\label{homomorphismf}
\end{equation}
where $F(M;\alpha_1\cup\alpha_2)$ is given by eq.(\ref{fmphi}),
$\alpha_i$ is a coloring on $\Sigma_i$ and the sum is taken over
all colorings on $\Sigma_2$.

By considering a cobordism $(M; \Sigma_2, \Sigma_1)$
instead of $(M; \Sigma_1, \Sigma_2)$
for the same triangulated manifold $M$,
\begin{equation}
F_{21}(\alpha_2)\,=\,\sum_{\alpha_1}F(M;\alpha_1\cup\alpha_2)\,\alpha_1
\quad : V_{-\Sigma_2}\rightarrow V_{-\Sigma_1}\, .
\end{equation}
If we identify the vector field $V_{\Sigma_i}$ with
its dual $V_{\Sigma_i}{}^\ast$,
$F_{21}$ is interpreted as 
a dual map of the linear map $F_{12}$. 
Thus, in this sense, 
\begin{lemma}
 $F_{21}{}^\ast \,=\, F_{12}$ .
\label{dualf}
\end{lemma}

Let $(M; \Sigma_1, \Sigma_3)$ be the composition of 
cobordisms $(M_1; \Sigma_1, \Sigma_2)$ and 
$(M_2; \Sigma_2, \Sigma_3)$, i.e., $M=M_1\cup M_2$. 
We fix a triangulation on $M$ and direction of arrows 
on 1-simplices. 
Then
\begin{lemma}
 $F_{23}\,\circ\, F_{12} \,=\, F_{13}$ .
\label{composition}
\end{lemma}
{\it Proof.}
\begin{eqnarray}
F_{23}\,\circ\,F_{12}(\alpha_1)\,&=&
\sum_{\alpha_2}F(M_1;\alpha_1\cup\alpha_2)\,F_{23}(\alpha_2)\nonumber\\
&=& \sum_{\alpha_2}F(M_1;\alpha_1\cup\alpha_2)
\sum_{\alpha_3}F(M_2;\alpha_2\cup\alpha_3)\,\alpha_3 
\label{composef}
\end{eqnarray}
where $\alpha_i$ is a coloring on $\Sigma_i$.
Let $N_i^1$, $N_i^2$ and $N_i$ be the numbers of $i$-simplices of
$M_1$, $M_2$ and $M$ respectively, and  
$v^j$ and $f^j$ be the numbers of 0-simplices and 
2-simplices of $\Sigma_j$.
Note that the following relation holds :
\begin{equation}
N_2=N_2^1+N_2^2-f^2, \quad 
N_0=N_0^1+N_0^2-v^2.
\end{equation}
Then, 
by using (\ref{fmphi}),
\begin{eqnarray}
F_{23}\,\circ\,F_{12}(\alpha_1)\,&=&
|X|^{-2N_2^1-N_0^1+f^1+f^2+(v^1+v^2)/2}\,
|X|^{-2N_2^2-N_0^2+f^2+f^3+(v^2+v^3)/2}
\nonumber\\
& & \hspace{1cm}\times 
\sum_{\alpha_2} \sum_{\alpha_3} 
\,\sum_{\{\phi\}|_{\alpha_1,\alpha_2,\alpha_3}} \,
\prod_{i=1}^{N_3^1}W_i{}^{\kappa_i}\prod_{j=1}^{N_3^2}W_j{}^{\kappa_j}
\nonumber\\
&=& 
|X|^{-2N_2-N_0+f^1+f^3+(v^1+v^3)/2}
\sum_{\{\phi\}|_{\alpha_1,\alpha_3}} 
\sum_{\alpha_3}\, 
\prod_{i=1}^{N_3}\,W_i{}^{\kappa_i}
\nonumber\\
&=& \sum_{\alpha_3}F(M;\alpha_1\cup\alpha_2\cup\alpha_3)\,\alpha_3\nonumber\\
&=& F_{13}(\alpha_1) .
\hspace{7cm}\rule[-1pt]{6pt}{6pt}
\end{eqnarray}
Note that a category of $\C$-modules whose 
object and morphism are $V_\Sigma$ and $F_{ij}$ : 
$\Sigma_i\rightarrow \Sigma_j$ 
is defined from the above lemma.

For a cobordism $(\Sigma\times I; \Sigma,\Sigma)$
between a triangulated manifold $\Sigma$,
we denote 
\begin{equation}
  F_{id_\Sigma}\, :\, V_{\Sigma}\longrightarrow V_{\Sigma}
\end{equation}
a homomorphism defined by eq.(\ref{homomorphismf}).
\begin{lemma}
\begin{equation}
  tr\; F_{id_\Sigma}\, =\, F(\Sigma\times S^1) . 
\label{tracef}
\end{equation}
\label{trace}
\end{lemma}
{\it Proof.}
\begin{eqnarray}
  tr\; F_{id_\Sigma}\,& =&\, 
      \sum_{\alpha_1=\alpha_2}F(M; \alpha_1\cup \alpha_2)\nonumber\\
    &=& |X|^{-2N_2-N_0+2f+v} 
                 \sum_{\{\phi\}|_{\alpha_1=\alpha_2}}
        \prod_{i=1}^{N_3} W_i{}^{\kappa_i}
\end{eqnarray}
where $v$ and $f$ is the number of 0-simplices and 2-simplices of $\Sigma$.
The last equation is equal to $F(\Sigma\times S^1)$ where 
the manifold $\Sigma\times S^1$ has 
$N_2-f$ 2-simplices  and $N_0-v$ 0-simplices. 
\rule[-1pt]{6pt}{6pt}

For an arbitrary cobordism $(M; \Sigma_1, \Sigma_2)$,
it is shown that 
\begin{equation}
  F_{12}\,=\, F_{12}\circ F_{id_{\Sigma_1}}
\end{equation}
is satisfied 
from proposition~\ref{invariantd} and lemma~\ref{composition}.
Thus
\begin{equation}
  Ker\,(F_{12})\; \supset\; Ker\,(F_{id_{\Sigma_1}}) .
  \label{id1}
\end{equation}
Furthermore, the equation
\begin{equation}
  Im(F_{12})\,=\, Im(F_{12})/Ker(F_{id_{\Sigma_2}}) 
  \label{id2}
\end{equation}
follows from 
\begin{equation}
  F_{12}\,=\,F_{id_{\Sigma_2}}\circ  F_{12} .
\end{equation}
By the equations (\ref{id1}) and (\ref{id2}), the map
\begin{equation}
  \Psi_{12}\,:\, Q_{\Sigma_1}\longrightarrow Q_{\Sigma_2}  
\end{equation}
is induced by a homomorphism
$F_{12} : V_{\Sigma_1}\rightarrow V_{\Sigma_2}$ 
where 
\begin{equation}
  Q_{\Sigma_i}\,=\, V_{\Sigma_i}/\,Ker(F_{id_{\Sigma_i}}) .
\end{equation}
Note that the map 
$\Psi_{id_\Sigma}:Q_{\Sigma}\rightarrow Q_{\Sigma}$
corresponding to 
$F_{id_\Sigma}$
is a monomorphism and
can be regarded as identity map on $Q_{\Sigma}$ 
if we choose the basis of $Q_{\Sigma}$ properly.
Thus the correspondence 
$\Sigma\rightarrow Q_\Sigma$
and $(M;\Sigma_1,\Sigma_2)\rightarrow \Psi_{12}$
forms a functor from the category of 
cobordisms with triangulation and arrows to the category of $\C$-modules.

Furthermore, by considering a map 
$$\Psi_{id_\Sigma} = \Psi_{\Sigma\Sigma'}\circ\Psi_{\Sigma'\Sigma}$$
for a manifold $M = \Sigma\times I$
with $\partial M = (-\Sigma)\cup \Sigma'$ where the
topology of the two triangulated 
 surfaces $\Sigma$ and $\Sigma'$
are the same, we can show the relation
\begin{equation}
  \dim\,Q_{\Sigma} = \dim\,Q_{\Sigma'} . 
\end{equation}
We identify $Q_{\Sigma}$ and $Q_{\Sigma'}$ by means of
the map 
$\Psi_{\Sigma\Sigma'} : Q_{\Sigma}\rightarrow Q_{\Sigma'}$.
Thus from this identification, the map
$\Psi_{\Sigma\Sigma'}: Q_{\Sigma}\rightarrow Q_{\Sigma'}$ 
is independent of the choice of 
triangulation and arrows on $Q_{\Sigma}$ and $Q_{\Sigma'}$.

The above arguments defines 
a functor from the 3-dimensional cobordisms of non-triangulated surfaces
to the category of $\C$-modules:
$\Sigma\rightarrow Q_\Sigma$
and $(M;\Sigma_1,\Sigma_2)\rightarrow \Psi_{12}$. 
Thus combining these results with 
lemma~\ref{dualf} and lemma~\ref{composition}, 
\begin{theo}
The functor defined as above is nothing but a functor of three-dimensional 
topological quantum field theory 
which satisfies the Atiyah's axioms~\cite{at}.
\end{theo}
The important consequence of the axioms of 
topological quantum field theory is 
\begin{equation}
  \dim Q_{\Sigma}\,=\, F(\Sigma\times S^1) ,
   \label{traceq}
\end{equation}
which is straightforward from lemma~\ref{trace}.

Remember that (\ref{zas3}) and (\ref{zas2s1}), i.e., for any choice of 
involutory Hopf algebra $A$,
$$ Z_A(S^2\times S^1) = 1, \qquad Z_A(S^3)=|X|^{-1}\,.$$ 
Then, we see from eq.(\ref{traceq})
\begin{equation}
\dim Q_{S^2}=1,
\end{equation}
which, with lemma~\ref{composition},  
leads to 
\begin{equation}
 Z_A(M)Z_A(S^3)=Z_A(M_1)Z_A(M_2)
\end{equation}
where $M=M_1 \# M_2$.

\section{Generalization}
In this section, we generalize the topological invariant 
$Z_A(M)$ ($=F(M)$)
to that of  
compact manifold $M$ with boundary, i.e.,
we give a topological invariant complex number to $M$.
\begin{figure}[tbhp]
    \begin{center}
      \leavevmode
       \epsfysize = 3.5cm 
       \epsfbox{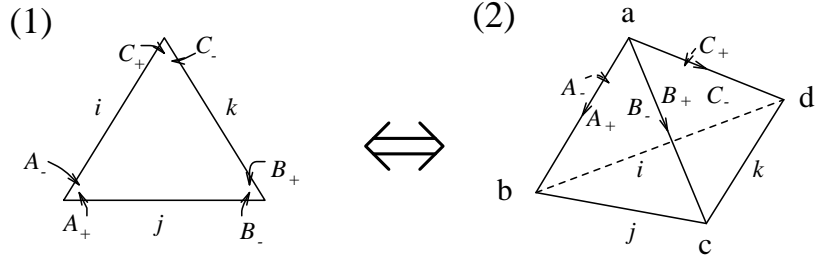}
    \end{center}
    \caption{A triangle with color. The weight $\tilde{W}_{F}{}^{\kappa}$
      defined for a colored triangle $F$ is given by the weight $W^{\kappa}$ 
      of the tetrahedron $T$ depicted in the right-hand side.
      For any $F$, we set $a>b$, $a>c$, $a>d$. }
\label{wtriangle}
\end{figure}
\begin{defi}
A vertex coloring on the boundary
$\partial M$ of a triangulated manifold $M$
is a map 
\begin{equation}
  \chi : \{(V_{i_0},E_{i_1}, F_{i_2})\, |\, i_k=1,\cdots,M_k (k=0,1,2),\, 
     E_{i_1}\cap F_{i_2} =E_{i_1},\,V_{i_0}\cap E_{i_1} =V_{i_0}
    \}\longrightarrow X
\label{verco}
\end{equation}
where $V_{i_0}$, $E_{i_1}$ and $F_{i_2}$
denote 0-, 1- and 2-simplex in $\partial M$ respectively 
and $M_k$ is the number of $k$-simplices in $\partial M$ 
(Fig.\ref{wtriangle}(1)).
\end{defi}
To give a vertex coloring is equivalent to give 
$6\times M_2$ ($= 4\times M_1$) symbols, which are elements of $X$, 
to all triples of eq.(\ref{verco}).
We assume that
all 1-simplices in $M$ are equipped with arrows
induced by an order of vertices as in section~\ref{secfm}.

Now we define a weight $\tilde{W}_{F}$ for a triangle 
$F\in \partial M$
with ordered vertices $b$, $c$ and $d$.
We temporarily put a new vertex $a$ in the outside of 
$F$ as $a\notin  M$ as $a>b$, $a>c$ and $a>d$ 
Then we make a tetrahedron $T$ of $a,b,c,d$
whose coloring is induced by that of $F$ as Fig.\ref{wtriangle}.
We define the weight
$ \tilde{W}_{F}{}^{\tilde{\kappa}}$
of $F$ as the weight $W^{\kappa}$
of the tetrahedron $T$ given above :
\begin{equation}
    \tilde{W}_{F}{}^{\tilde{\kappa}}
        \tuple {i_+,i_-} {j_+,j_-} {k_+,k_-} {A_+,A_-} 
      {B_+,B_-} {C_+,C_-} \equiv
    W^{\kappa} \tuple {i_+,i_-} {j_+,j_-} {k_+,k_-} {A_+,A_-} 
      {B_+,B_-} {C_+,C_-} 
\end{equation}
where $i_{\pm}$, $j_{\pm}$ and $k_{\pm}$ are colors on edges of 
a triangle $F$ and $A_{\pm}$, $B_{\pm}$ and $C_{\pm}$ 
are colors on vertices on $F$ (Fig.\ref{wtriangle}).
{}From the assignment of the weight 
$\tilde{W}_{F_i}{}^{\tilde{\kappa}_i}$ 
for each triangle $F_i$ which belongs to the boundary
$\partial M$ of $M$ in addition to that of $W_i{}^{\kappa_i}$ 
for each 3-simplices,
the following quantity is defined : 
\begin{equation}
  \tilde{F}(M) = |X|^{-2N_2-N_0-2M_1}
    \sum_{\phi} 
    \prod_{i=1}^{N_3}W_i{}^{\kappa_i}
    \sum_{\chi} \prod_{j=1}{}^{M_2}\tilde{W}_{F_j}{}^{\tilde{\kappa}_j}
    \in \C .
\label{tildefm}
\end{equation}
\begin{theo}
 $\tilde{F}(M)$ is independent of 
  the  direction of arrows on  1-simplices of $M$
 and any local topology preserving deformation 
 of $M$. 
\end{theo}
{\it Proof.}

Note that $\tilde{F}(M)$ is equal to $Z_A(M)$ in the case of 
$M$ being a closed manifold.
Therefore 
the invariance of $\tilde{F}(M)$ 
under the local deformation of $M$ which does not change 
the triangulation of $\partial M$ is verified by 
applying Prop.\ref{invariantd}. 
Furthermore,  
the independence of direction of arrows 
on 1-simplices in $M$ 
is easily shown by the relation
\begin{equation}
   \Delta^{x_1x_2\cdots x_n}=
   \Delta^{y_n y_{n-1}\cdots y_1}
   S^{x_1}{}_{y_1}S^{x_2}{}_{y_2}\cdots S^{x_n}{}_{y_n}.
\end{equation}

Thus we have only to show the invariance of 
$\tilde{F}(M)$ under the local deformation which changes   
the triangulation on the boundary.  
Remember that all topology preserving moves
of a triangulated surface is generated by finite 
sequence of $(2,2)$-moves, $(3,1)$-moves
and $(1,3)$-moves~\cite{gv}.
In the case of a triangulated 
surface that is a boundary of a triangulated 3-manifold $M$,
the moves are induced by adding a new 3-simplex to the boundary.
We provide two types of moves.
One is 
a `$(2,2)$-move' on $\partial M$ induced by the 
addition of two new triangles, and thus a new tetrahedron, 
 to $M$ as Fig.\ref{move}(1).
The other is a `$(3,1)$-move'
induced by adding a triangle 
to form a new tetrahedron on $M$
as Fig.\ref{move}(2).
Note that under these moves the numbers 
$N_2$, $N_0$ and $M_1$
are changed as 
\begin{eqnarray} 
{\rm `(2,2)-move'} & &\quad (N_2,N_0,M_1)\longrightarrow (N_2+2,N_0,M_1) , 
\label{22n}\\
{\rm `(3,1)-move'} & &\quad (N_2,N_0,M_1)\longrightarrow (N_2+1,N_0,M_1-3) .
\label{31n}
\end{eqnarray} 
These two types of moves 
is sufficient for all topology preserving deformation on 
$\partial M$
because $(1,3)$-moves are generated by 
the inverse operation of the `$(3,1)$-move' described above, i.e.,
removing a triangle $F$ from $\partial M$ 
after reforming a triangulation of $M/\partial M$
so that every edge of $F$ is a boundary of at least three triangles.
The invariance of $\tilde{F}(M)$ 
under `$(2,2)$-moves' and `$(3,1)$-moves' 
is verified explicitly 
by using the 
relations 
(\ref{32move}) and (\ref{32move2}) 
which describe the 
`$(2,2)$-move' and the `$(3,1)$-move' respectively as follows :
\begin{description}
\item{`$(2,2)$-move' :}
 \begin{eqnarray}
  &&\hspace{-5mm} |X|^{4}
 \tilde{W}^+ \tuple {i_1,i_2} {j_1,j_2} {k_1,k_3} {B_3,B_2} 
      {D_1,D_3} {C_1,C_2} 
 \tilde{W}^- \tuple {l_1,l_2} {m_1,m_2} {k_3,k_2} {D_3,D_2} 
      {B_1,B_3} {A_1,A_2} 
\nonumber\\
&=& 
 \tilde{W}^+ \tuple {n_1,n_2} {j_3,j_2} {l_3,l_2} {B_1,B_2} 
      {A_1,A_3} {C_1,C_3} 
\nonumber\\
&&\hspace{1cm}\times
 W^+ \tuple {i_3,i_2} {j_1,j_3} {k_1,k_2} {l_1,l_3} 
      {m_3,m_2} {n_2,n_3} 
 \tilde{W}^- \tuple {n_3,n_1} {m_1,m_3} {i_1,i_3} {D_1,D_2} 
      {C_3,C_2} {A_3,A_2} 
 \end{eqnarray}
\begin{figure}[tbhp]
    \begin{center}
      \leavevmode
       \epsfysize = 5.5cm 
       \epsfbox{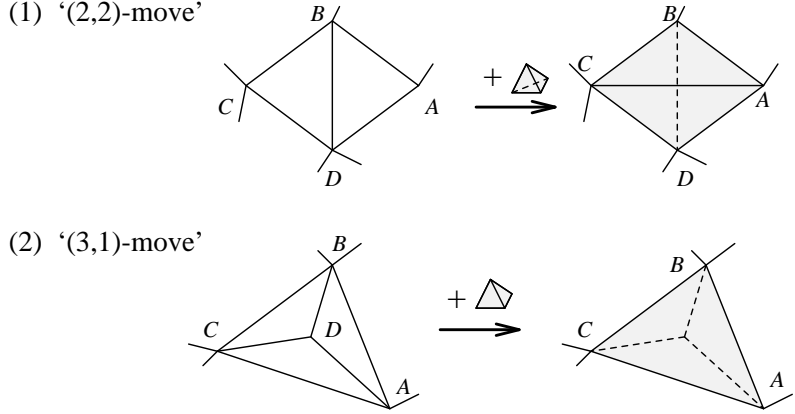}
    \end{center}
    \caption{`$(2,2)$-move' and `$(3,1)$-move' on the
             boundary $\partial M$ of $M$.
    The addition of a tetrahedron to $\partial M$ 
    generates the local move, 
(1) or (2), on $\partial M$.}
\label{move}
\end{figure}
\item{`$(3,1)$-move' :}
\begin{eqnarray}
&&\hspace{-5mm} |X|^{-4}
 \tilde{W}^+ \tuple {t_3,t_2} {u_1,u_3} {j_1,j_2} 
            {C_1,C_3} {A_3,A_2} {D_1,D_2} 
\nonumber\\
&&\hspace{1cm}\times
 \tilde{W}^+ \tuple {s_3,s_2} {t_1,t_3} {k_1,k_2} 
            {A_1,A_3} {B_3,B_2} {D_2,D_3} 
 \tilde{W}^- \tuple {s_1,s_3} {i_1,i_2} {u_3,u_2} 
            {C_3,C_2} {D_3,D_1} {B_1,B_3} 
\nonumber\\
&=& 
 \tilde{W}^+ \tuple {i_3,i_2} {j_3,j_2} {k_3,k_2} 
            {A_1,A_2} {B_1,B_2} {C_1,C_2} 
 W^+ \tuple {t_1,t_2} {u_1,u_2} {j_1,j_3} 
            {i_1,i_3} {k_1,k_3} {s_1,s_2} \quad .
\end{eqnarray}
\end{description}
These relations with (\ref{22n}) and (\ref{31n})
ensure that the value 
$\tilde{F}(M)$ of eq.(\ref{tildefm})
does not change under 
`$(2,2)$-move' and `$(3,1)$-move.'
Since all topology preserving deformations of $M$ is generated
by these moves and their inverses together with 
the local deformations of $M$ which do not change 
the triangulation of $\partial M$, 
the proof is completed.
\rule[-1pt]{6pt}{6pt}

The following formula is verified from the definition :
\begin{equation}
   \tilde{F}(M\setminus\,{\rm Int}\, D^3)= |X|\tilde{F}(M)
\end{equation}
where $D^3$ denotes the three-dimensional ball.

\section{Concluding Remarks}
In this paper,
we generalized the Chung-Fukuma-Shapere invariant $Z_A(M)$
of closed three-dimensional manifold $M$ 
to the functor $\Psi_{ij}$ of a topological quantum field theory
which satisfies Atiyah's axioms.
We also generalized $Z_A(M)$ to the invariant $\tilde{F}(M)$
of a compact manifold $M$ with boundary.

The crucial point of defining the functor and $\tilde{F}(M)$
for a triangulated manifold $M$
is to give the weight $W_i{}^{\kappa_i}$ to each
tetrahedron $T_i$.
The weight in this theory plays the same role as that of  
quantum $6j$-symbols in the Turaev-Viro theory.
After defining $W_i{}^{\kappa_i}$,
it is straightforward to define the functor and $\tilde{F}(M)$ 
by referring to the Turaev-Viro theory and its generalization
in ref.\cite{kms}.
Note that the weight $W_i{}^{\kappa_i}$ is defined for 
colors on every pair of adjacent $edge$ and $face$ of $T_i$
with arrows.
On the other hand,
quantum $6j$-symbol is defined for colors on edges 
without arrows.

Finally, we give two remarks;
we can define the functor and $\tilde{F}(M)$
for any cell complex which $Z_A(M)$ is defined on,
though we do not give the definition; 
Kuperberg generalized his invariant, which is equivalent to $Z_A(M)$,
for non-involutory Hopf algebras \cite{ku2}.

\subparagraph{Acknowledgments}
The author thanks S.~Higuchi for valuable discussions.
She is grateful to S.~Higuchi and N.~Sakai for 
careful reading of the manuscript.
This work is supported  by
Grant-in-Aid for Scientific Research from the Ministry of
Education, Science and Culture.

\end{document}